\documentclass[aps,11pt]{revtex4-1}

\usepackage{graphicx}
\usepackage{amsfonts}
\usepackage{amsmath}
\usepackage{mathrsfs}
\usepackage{eucal}
\usepackage{appendix}
\usepackage[ruled]{algorithm2e}
\usepackage{natbib}

\newcommand\bR{\mathbb{R}}

\newcommand\mD{\mathcal{D}}

\newcommand\mF{\mathcal{F}}

\newcommand\mI{\mathcal{I}}

\newcommand\mM{\mathcal{M}}

\newcommand\mA{\mathcal{A}}
\newcommand\mS{\mathcal{S}}

\newcommand\bC{\mathbf{C}}

\def\o{\omega}

\begin{document}

\title{Information-related complexity: a problem-oriented approach}
\author{E. Perevalov}
\email[E-mail: ]{eup2@lehigh.edu}
\author{D. Grace}
\email[E-mail: ]{dpg3@lehigh.edu}
\affiliation{Lehigh University\\ Bethlehem, PA}

\begin{abstract}
A general notion of information-related complexity applicable to both natural and man-made systems is proposed. The overall approach is to explicitly consider a rational agent performing a certain task with a quantifiable degree of success. The complexity is defined as the minimum (quasi-)quantity of information that's necessary to complete the task to the given extent -- measured by the corresponding loss. The complexity so defined is shown to generalize the existing notion of statistical complexity when the system in question can be described by a discrete-time stochastic process. The proposed definition also applies, in particular, to optimization and decision making problems under uncertainty in which case it gives the agent a useful measure of the problem's ``susceptibility'' to additional information and allows for an estimation of the potential value of the latter.
\end{abstract}



\maketitle

\section{\label{s:intro}Introduction}
The general field of complex systems science is relatively new and has been developing rather rapidly in the last three decades. One of the central notions it revolves around is -- understandably -- that of complexity. The desire to define complexity before attempting to study it is indeed rather natural and logical in this context. When the field of complex system science began taking its shape several notions of complexity were already in use. The most notable of them was the Kolmogorov descriptional complexity \cite{KOLMOGOROV:1958,KOLMOGOROV:1968} defined as a shortest length of a computer program needed to reproduce the given object (in the simplest case represented by a string of symbols from a given alphabet). It was quickly noted, however, that the Kolmogorov complexity is rather a measure of {\it randomness} as opposed to complexity understood as the amount of appropriately defined ``internal structure'' possessed by the object \cite{hogg85}. It was also realized that a proper measure of complexity should be probabilistic since complexity as it is encountered in natural sciences and engineering is inherently linked to some sort of {\it uncertainty}. Following these realizations, several useful measures of complexity were developed, including set complexity, true measure complexity, effective measure complexity \cite{grassberger86}, logical depth \cite{bennett85,bennett86}, statistical complexity \cite{crutchfield89}, sophistication \cite{koppel87}. Among these measures, statistical complexity was the measure foundations and applications of which were later developed in most detail (see \cite{shalizi-crutch01}). For this reason, we will use it later as a ``benchmark'' complexity measure for systems that can be described by discrete-time stochastic processes.

The main goal of the present work is to put the problem of complexity measures in an ``engineering'' problem-oriented context and, in particular, to demonstrate that a generalization of statistical complexity measure arises naturally when the proposed ``engineering'' complexity notion is applied to the case of a prediction task for discrete-time stochastic processes. The central figure of the proposed approach is an {\it agent} whose goal is to perform a certain {\it task}. Typical examples of tasks are control, decision making, optimization, prediction. We will also refer to the such tasks as {\it problems}. From a general practical point of view, a task can be made complex by several of its aspects. It can be, for example, computationally complex (like a large scale deterministic optimization problem). It can be operationally complex (like that of building a complicated object given full instructions). We will not deal with such complexity types. Instead, we concentrate on {\it information}-related complexity, the measure of which is an appropriately defined quantity of some relevant information needed for completing the task with a given (also appropriately measured) degree of success. This implies that the task is initially associated with some degree of uncertainty that can be fully or partially removed by acquiring the appropriate information.

The rest of the paper is organized as follows. In Section~\ref{s:general}, the general definition of task complexity is given. Section~\ref{s:opt} applies considers the optimization under uncertainty (stochastic optimization) as a task and derives the quasi-information complexity for optimization. Section~\ref{s:stochastic} considers the task of prediction of stochastic processes output and makes a connection to statistical complexity introduced in \cite{crutchfield89} which turns out to be just the task complexity taken at zero value of the argument. Finally, Section~\ref{s:conclusion} summarizes the contributions of the present work and presents a brief informal discussion. 

\section{\label{s:general}General formulation: information-related task complexity}
An agent is trying to accomplish a certain task and can, in principle, be helped by appropriate additional information. The quality (or degree) of the task accomplishment is assumed to be quantifiable by a real number and thus can be compared between any two instances of the task accomplishment. Respectively, a notion of real-valued {\it loss} can be defined as (the absolute value of) the difference between the degree of the task accomplishment in the given instance and the best possible one. The latter has to be defined appropriately given the specific task at hand as examples given later in the paper illustrate.

Besides loss, any information $\mI$ that the agent can acquire is also assumed to possess some real-valued measure $I(\mI)$ of its effective quantity which we will refer to as {\it quasi-quantity} to distinguish it from measures of ``pure'' quantity such as Shannon entropy. The appropriate form of quasi-quantity measure $I(\mI)$ has to be determined by considering the specific nature of the task in hand to appropriately reflect its essential features. In particular, the quasi-quantity measure may coincide with a traditional ``pure'' quantity measure for some tasks. The agent is also assumed to be capable of making optimal use of the acquired information for the purpose of minimizing the loss. Given the latter assumption, we can introduce the notion of the loss $L(\mI)$ of the specific information $\mI$ available to the agent.

It is natural to assume that, if the measure of information quantity adapted in the given context is adequate, then smaller loss values can in general be obtained for larger values of the information quasi-quantity. Thus, to reduce the loss more, larger quantities of information would in general be needed. On the other hand, it is clear that, in general, different ``pieces'' of information of equal quasi-quantity would lead to different values of loss reduction. Therefore, if loss reduction at a minimum information ``cost'' is the agent's goal, it makes sense to search for the specific information that would allow the agent to achieve that goal. It is also rather natural to use the initial value of the loss $L_0\equiv L(\emptyset)$ that the agent can obtain in the absence of any (additional) information as a reference point for the reduced loss. Specifically, one can use the ratio $\gamma=\frac{L(\mI)}{L_0}$ as a natural dimensionless measure of the achieved loss reduction. The lowest information quasi-quantity needed to achieve such a loss reduction can be associated with the proper {\it quasi-information complexity} $C_{\gamma}$ of the task in question. This leads us to the following definition.
\begin{equation}
C_{\gamma}=\min_{\{\mI : L(\mI)\le \gamma L_0\}} I(\mI).
\label{eq:Cgamma}
\end{equation}
Note that the minimization in (\ref{eq:Cgamma}) is performed over all possible ``kinds'', or ``pieces'' of information that -- if used optimally by the agent -- would yield the value of loss not exceeding the given fraction $\gamma$ of the original value $L_0$.
Note also that the specific interpretations of the loss $L(\cdot)$ and quantity $I(\cdot)$ are intentionally left open by the definition (\ref{eq:Cgamma}). The understanding is that they have to be made more concrete by taking the ``nature'' and the details of the system in question into account. In the next two sections, we consider two more specific examples that illustrate the use of the proposed definition.

\section{\label{s:opt}Optimization}
Consider a typical ``engineering'' case of optimization under uncertainty, with the expected value of the objective function being the main optimization criterion (stochastic optimization formulation). The overall task can be concisely written as follows.
\begin{equation}
 \label{eq:gen_stoch}
 \mbox{min}_{x\in X} \int_{\Omega} f(\omega, x) P(d\o),
 \end{equation}
  where $X\subset \mD$ is the set of all feasible solutions, i.e. the set satisfying all (deterministic) constraints that may be present in the problem formulation (with $\mD$ being the space to which all solutions belong  like a suitable Euclidean space);  $\Omega$ is the base space of possible values of input data parameters that are not known with certainty often referred to as a {\it parameter space}; $P$ is a probability measure on ($\Omega,\mF)$  (where $\mF$ is a suitable sigma-algebra) that describes the initial state of information available to the agent; $f$: $\Omega\times \mD\rightarrow \overline \bR$ is a function integrable on $\Omega$ for each $x\in X$ describing the specific optimization objective.

The value of the initial loss $L_0$ then becomes simply
\begin{equation}
L_0=L(P)=\int_{\Omega} (f(\o,x^*_P) - f(\o,x^*_{\o})) P(d\o)
\label{eq:L_0}
\end{equation}
here $x^*_P$ is a solution of (\ref{eq:gen_stoch}) and $x^*_{\o}$ is a solution of $\mbox{min}_{x\in X} f(\o,x)$ for the given $\o$ of the parameter space $\Omega$.

As it has been argued in \cite{part1} and \cite{part2}, in the general context of problem (\ref{eq:gen_stoch}), additional information can be obtained by the agent by means of querying {\it information sources}. This can be accomplished by the agent via formulating and asking {\it questions} and receiving corresponding {\it answers} from the source(s). It has been further argued that the appropriate meaning of information quasi-quantity  is supplied by {\it pseudoenergy} that provides real-valued measure for both {\it question difficulty} and {\it answer depth} functionals. The reason pseudoenergy is appropriate in the given context is that it takes into account the source's {\it knowledge structure} and therefore allows the agent to correctly select optimal ``pieces'' of information to maximize their effect on the given problem.

The role of ``pieces'' of information themselves is played by questions which can be associated, following the earlier developments of \cite{cox1946,cox1961,cox1979} and \cite{knuth05,knuth08}, with {\it partitions} of the problem parameter space $\Omega$. The corresponding quasi-quantity is then given by the question difficulty functional $G(\Omega,\bC,P)$ where the partition $\bC=\{C_1,\dotsc, C_r\}$ is a collection of subsets of $\Omega$ such that $C_i\cap C_j=\emptyset$ for $i\ne j$ and $\cup_{i=1}^r C_i=\Omega$. The loss $L(\bC)$ corresponding to question $\bC$ can be interpreted as the minimum loss that an optimizing agent can obtain upon reception of a perfectly accurate answer to question $\bC$ can be easily seen to take the form
\begin{equation}
L(\bC,P)=\sum_{j=1}^r P(C_j)\int_{C_j}(f(\o, x_{P_j}^*)-f(\omega,x^*_{\o})) dP_{C_j}(\o),
\label{eq:L(bC)}
\end{equation}
where $P_{C_j}$ is the conditional measure corresponding to the subset $C_j\in \bC$. It is straightforward to show that $L(\bC)\le L_0$. Correspondingly, the informational complexity $C_{\gamma}$ of the optimization task (\ref{eq:gen_stoch}) takes the form
\begin{equation}
C_{\gamma}=\min_{\{\bC : L(\bC,P)\le \gamma L(P)\}} G(\Omega,\bC,P),
\label{eq:Cgamma-opt}
\end{equation}
where minimization in (\ref{eq:Cgamma-opt}) is over all possible partitions of $\Omega$. The form of informational complexity (\ref{eq:Cgamma-opt}) assumes a specific source knowledge structure which is encoded in the form of question difficulty functional $G(\cdot)$. In particular, as was shown in \cite{part1}, in an isotropic (but in general non-uniform) case, the functional $G(\cdot)$ depends besides the ``initial'' probability measure $P$, on an integrable function $u$: $\Omega\rightarrow \bR_+$ on the parameter space $\Omega$:
\begin{equation*}
G(\Omega, \bC, P)=\frac{\sum_{j=1}^r u(C_j)P(C_j)\log \frac{1}{P(C_j)}}{\sum_{j=1}^r P(C_j)},
\end{equation*}
where $u(C_j)=\frac{\int_{C_j}u(\o)\,dP(\o)}{P(C_j)}$.
If the knowledge structure is unknown, it appears reasonable to assume the most symmetric case (uniform and isotropic) in order to obtain a ``benchmark'' (or ``universal'') task complexity. In this case, the question difficulty reduces to Shannon entropy $H(P(\bC))\equiv H(P(C_1),\dotsc, P(C_r))$ of the discrete probability distribution induced by the partition $\bC$, and the informational task complexity $C_{\gamma}$ becomes simply
\begin{equation}
C_{\gamma}=\min_{\{\bC : L(\bC,P)\le \gamma L(P)\}} H(P(\bC)).
\label{eq:Cgamma-opt-uni}
\end{equation}

\subsection{Example}
\begin{figure}
\includegraphics[scale=0.8]{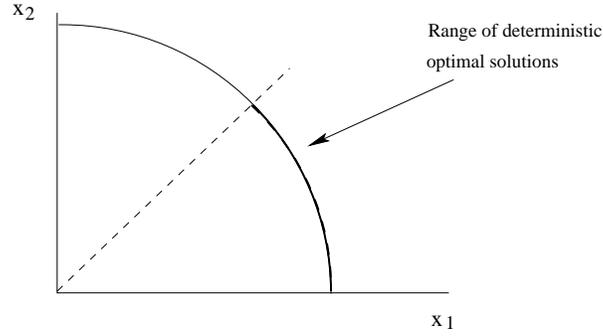}
\caption{\label{f:opt-ex1}Example 1.}
\end{figure}

\begin{figure}
\includegraphics[scale=0.4]{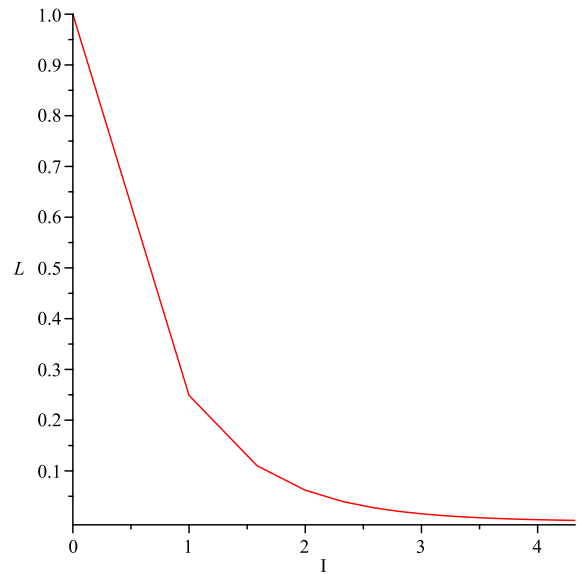}
\caption{\label{f:loss-SO} Loss vs. information quasi-quantity for example 1.}
\end{figure}

Consider the stochastic optimization problem of the form
\begin{equation*}
\begin{aligned}
& {\text{maximize}}
& & x_1+\omega x_2  \\
& \text{subject to}
& & x_1^2+x_2^2\le 1  \\
& & & x_1\ge 0,\; x_2\ge 0\\
\end{aligned}
\end{equation*}
where the only uncertain parameter is $\o$ which has a uniform distribution on the interval $\Omega=[0,1]$. The situation is depicted in Fig.~\ref{f:opt-ex1} and is referred to as Example~1 in what follows. It is straightforward to show that for a partition $\bC$ of $\Omega$ into $r$ subsets of equal measure, the loss (\ref{eq:L(bC)}) takes the form
\begin{equation*}
L(\bC,P)=\int_0^1 \sqrt{1+\o^2}\, d\o - \frac{1}{r}\sum_{i=1}^r \sqrt{1+\left(\frac{2i-1}{2r} \right)^2}.
\end{equation*}
On the other hand, if one uses (in the absence of information about possible information sources and their knowledge structure) the uniform isotropic model, the difficulty functional reduces to Shannon entropy and   $G(\Omega,\bC,P)=H(P(\bC))=\log r$. The relationship between the information quasi-quantity $H(P(\bC)$ and the corresponding loss (rescaled so that  $L_0=0$)  is shown in Fig.~\ref{f:loss-SO}. The task complexity $C_{\gamma}$ for some values of the parameter $\gamma$ are shown in Table~\ref{t:ex1-O}. In particular, the ``full'' task complexity $C_0$ for this example is infinite.

\begin{table}
\caption{\label{t:ex1-O} Quasi-information complexity $C_{\gamma}$ for some values of $\gamma$ for Example~1.}
\begin{ruledtabular}
\begin{tabular}{llllll}
        $\gamma$ & 0.2 & 0.1 & 0.05 & 0.01 & 0 \\
        $C_{\gamma}$ & 1.2 & 1.7 & 2.2 & 3.8 & $\infty$ \\
\end{tabular}
\end{ruledtabular}
\end{table}

\begin{figure}
\includegraphics[scale=0.8]{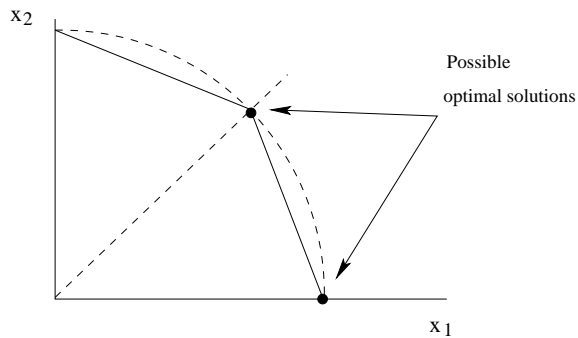}
\caption{\label{f:opt-ex2}Example 2.}
\end{figure}

\begin{figure}
\includegraphics[scale=0.4]{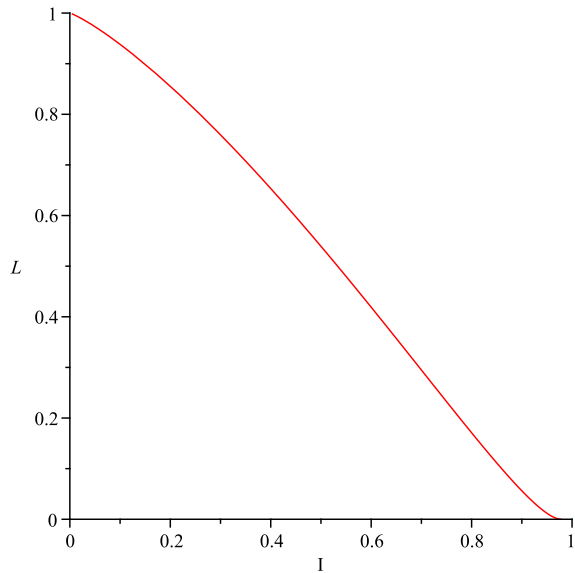}
\caption{\label{f:loss-SO1} Loss vs. information quasi-quantity for example 2.}
\end{figure}

Suppose now that the parameter space $\Omega$ and the initial measure $P$ are the same but the task is somewhat different. Specifically, the stochastic optimization problem has the linear form (see Fig.~\ref{f:opt-ex2} for an illustration). We refer to this example as Example~2.
\begin{equation*}
\begin{aligned}
& {\text{maximize}}
& & x_1+\o x_2  \\
& \text{subject to}
& & x_1+(\sqrt 2 -1)x_2 \le 1  \\
& & &(\sqrt 2 -1)x_1+x_2 \le 1  \\
& & & x_1\ge 0,\; x_2\ge 0\\
\end{aligned}
\end{equation*}
In this case, as it is easy to see, it is sufficient to consider questions corresponding to partitions of $\Omega$ into two subsets only, in order to fully get rid of the agent uncertainty with respect to the optimal solution -- assuming that perfectly accurate answers to such questions are available.
Specifically, consider partitions of $\Omega$ of the form
${\bf C}=\{[0,a], (a,1]\}$ where $0\le a \le \omega_c=\sqrt 2 -1$. Then the loss (\ref{eq:L(bC)})
takes the form
\begin{equation*}
L(\bC,P)=\frac{\sqrt 2 -1}{\sqrt 2}(\omega_c -a)-\frac{1}{2\sqrt 2}(\omega_c^2-a^2),
\end{equation*}
and the information quasi-quantity becomes
$$G(\Omega,\bC,P)=H(P(\bC))= H(a)=-a\log a -(1-a)\log (1-a).$$
The relationship between them is shown in Fig.~\ref{f:loss-SO1}. The task complexities for some values
of $\gamma$ are shown in Table~\ref{t:ex2-O}. One can see that the ``full'' complexity $C_0$ for this example is finite.

\begin{table}
\caption{\label{t:ex2-O} Quasi-information complexity $C_{\gamma}$ for some values of $\gamma$ for Example~2.}
\begin{ruledtabular}
\begin{tabular}{llllll}
        $\gamma$ & 0.2 & 0.1 & 0.05 & 0.01 & 0 \\
        $C_{\gamma}$ & 0.77 & 0.85 & 0.90 & 0.95 & 0.98 \\
\end{tabular}
\end{ruledtabular}
\end{table}

Let us now make some observations (and careful generalizations).
\begin{itemize}
\item  Even if the ``uncertainty structure'' (the parameter space $\Omega$ and the measure $P$) is precisely the same, the task complexities may be different (and even very different) if the agent is solving different problems. In the two examples just discussed, for instance, the complexity $C_0$ is infinite for Example~1 while it is and finite (and rather small) for Example~2.
\item The ``full'' complexity $C_0$ typically does not tell the whole story about the ``information content'' of a problem.
\item The whole curve $C_{\gamma}$ is in principle needed to obtain a picture of the problem's ``information susceptibility'', or, equivalently, the ability of additional information to improve the solution quality. 
\item The knowledge of the curve $C_{\gamma}$, coupled with that of available additional information sources and their knowledge structure (see \cite{part1,part2} for a discussion of the latter), allows the agent to evaluate the effectiveness of additional information for the particular problem and its corresponding value.
\end{itemize}

\section{\label{s:stochastic}Stochastic processes and statistical complexity}
\begin{figure}
\includegraphics[scale=0.8]{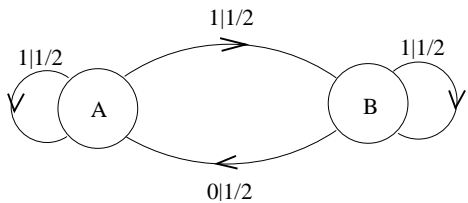}
\caption{\label{f:ab-hmm}Generation of observed symbols.}
\end{figure}

Consider now an application from a different domain: the case of discrete-time stochastic processes. To make the discussion more concrete, let us consider a specific simple example (referred to as Example~A in what follows) which is borrowed from \cite{crutchfield94}. Suppose the unobserved symbols $A$ and $B$ are generated randomly with equal probability. Symbols 0 and 1 -- that are observed -- correspond to pairs of $A$ and $B$ as follows:
$\{AA, AB, BB\}\rightarrow 1$, $\{BA\}\rightarrow 0$ (see Fig.~\ref{f:ab-hmm} for an illustration). For example, if $AABAB$ was generated, the observer will see 1101.

\begin{figure}
\includegraphics[scale=0.8]{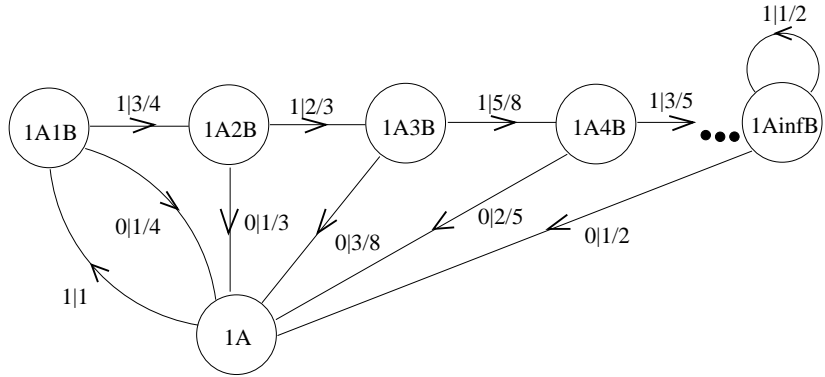}
\caption{\label{f:eps-mach-exact} Exact $\epsilon$-machine for Example~A.}
\end{figure}

The concept of $\epsilon$-machines was introduced in \cite{crutchfield89} and fully developed in \cite{shalizi-crutch01} to describe the intuitive (and practical) notion of complexity as the amount of underlying ``structure'' in the system. $\epsilon$-machines consist of casual states and transitions between them (see \cite{shalizi-crutch01} for a detailed exposition). Each time a transition happens, an observed symbols is output. They can be thought of as functions on Markov chains. Causal states are equivalence classes of previous histories: given two histories belonging to the same causal state $S$ the conditional distribution of future string of bits is the same. Thus $\epsilon$-machines are {\it deterministic} meaning that given a causal state and an output symbol, the next causal state is uniquely determined. $\epsilon$-machines can in principle be reconstructed from binary tree representation of the system output. Considering longer symbol sequences (words) allows for distinguishing more distinct causal states (which in turn requires longer output sample sizes). It can be shown that, for our example, the full $\epsilon$-machine has the form shown in Fig.~\ref{f:eps-mach-exact}. The transition probability for this $\epsilon$-machine have the form $\Pr(1AiB\rightarrow 1A(i+1)B)=\frac{i+2}{2(i+1)}$, $\Pr(1AiB\rightarrow 1A)=\frac{i}{2(i+1)}$, and $\Pr(1A\rightarrow 1A1B)=1$. Any transition to the state $1A$ outputs a symbol 0; all other transitions output 1.

Given an $\epsilon$-machine $\mM$ for the system, the {\it statistical complexity} $C_{\mu}$ is defined as the minimum expected amount of information that needs to be kept track of for optimal prediction of the next bit (which in general is imperfect):
\begin{equation}
C_{\mu}(\mM)= -\sum_{S\in \mM} P(S)\log P(S),
\label{eq:C_mu}
\end{equation}
where $P(S)$ is the (steady state) probability of the causal state $S$.
The {\it topological complexity} is a related quantity that reflects just the number of causal states in the $\epsilon$-machine:
$$C_{\tau}=\log|\mM| . $$
For our example, it is easy to see that $P(1AiB)=\frac{i+1}{2^{i+2}}$, $i=1,2\dotsc $ and $P(1A)=1/4$. Thus $C_{\mu}(\mM)=2.71$.  Note that the topological complexity $C_{\tau}=\log |\mM|$ of this system is infinite.

The {\it entropy rate} for an $\epsilon$-machine (and the corresponding dynamic system) can be found as
\begin{equation}
h_{\mu}(\mM)=-\sum_{S\in \mM} P(S) \sum_{s\in \mA} P_{\mM}(s|S)\log P_{\mM}(s|S),
\label{eq:h_mu}
\end{equation}
where $\mA$ is the alphabet of output symbols ($\mA=\{0,1\}$ in this example).
The entropy rate can be thought of as the measure of uncertainty per symbol that cannot be removed by finding the inherent structure of the system. For our example, the entropy rate is found to be equal to 0.678. We will also need the notion of {\it relative entropy rate} $h_{\mu}(\mM|\mM')$ of
$\epsilon$-machine $\mM$ with respect to another $\epsilon$-machine \footnote{Here an $\epsilon$-machine is understood as simply a function on a Markov chain, without a direct connection to any dynamic system or stochastic process.} $\mM'$. Let $g$: $\mM'\rightarrow \mM$ be a map between states of these two $\epsilon$-machines. Then the relative entropy rate $h_{\mu}(\mM|\mM')$ is defined as
\begin{equation*}
h_{\mu}(\mM|\mM')= -\sum_{S'\in \mM'}P_{\mM'}(S')\sum_{s\in \mA} P_{\mM'}(s|S')\log P_{\mM}(s|g(S')).
\end{equation*}
It is straightforward to show that
\begin{equation}
h_{\mu}(\mM|\mM')=h_{\mu}(\mM')+\sum_{S'\in \mM'}P_{\mM'}(S')KL(P_{\mM'}(S')||P_{\mM}(S')),
\label{eq:hrel}
\end{equation}
where $KL(P_{\mM'}(S')||P_{\mM}(S'))=\sum_{s\in \mA}P_{\mM'}(s|S')\log \frac{P_{\mM'}(s|S')}{P_{\mM}(s|g(S'))}$ is the corresponding Kullback-Liebler distance.

To make a link between task complexity and statistical complexity, we need to decide on the suitable task. It appears reasonable, since in this case we want to study the complexity of our system ``in its own right'', the logical choice for the task is that of suitably defined {\it prediction}. We also need to define the corresponding loss in a logical way. Several choices are in principle possible here, with {\it quadratic} and {\it logarithmic} loss being the two popular ones in prediction algorithms literature. We will use the latter of the two. Specifically, we will use the following single bit loss function
$$l(s,\sigma)=\begin{cases}
-\log \sigma, & \text{if}\; s=1 \\
-\log(1-\sigma), & \text{if}\; s=0
\end{cases},$$
where $0\le \sigma\le 1$ is the prediction. It's easy to see that if $p=\Pr(s=1)$ then
\begin{equation}
\min_{\sigma}E[l(s,\sigma)]=H(p)
\label{eq:min-loss}
\end{equation}
with the minimum being achieved for $\sigma=p$. Furthermore, one can show that
\begin{equation}
E[l(s,\sigma)]=H(p)+KL(p||\sigma),
\label{eq:gen-loss}
\end{equation}
where $KL(p||\sigma)=p\log\frac{p}{\sigma}+(1-p)\log\frac{1-p}{1-\sigma}$ is the Kullback-Liebler distance between distributions $(p,1-p)$ and $(\sigma,1-\sigma)$.

Comparing (\ref{eq:min-loss}) and (\ref{eq:h_mu}) we can see that the entropy rate for the given $\epsilon$-machine represents the minimum possible long-term (i.e. expected) logarithmic prediction loss. In other words, this is the minimum prediction loss that can be obtained by an optimally acting agent in possession of the complete information about the system. It is reasonable to identify the information $\mI$ from the general definition (\ref{eq:Cgamma}) with some $\epsilon$-machine \footnote{Strictly speaking, according to definitions of \cite{shalizi-crutch01}, the ``partial $\epsilon$-machines'' used here to describe incomplete information are not $\epsilon$-machines in the proper sense since their states are not {\it causal states} (i.e. equivalence classes of past histories characterized by the same distributions of futures, or {\it morphs}) , but rather simply {\it effective states} (i.e. equivalence classes of past histories that in general do not possess that property). Nevertheless, we refer to them as (``partial'') $\epsilon$-machines just to avoid introducing additional notions.} describing the system's output. Depending on the particular $\epsilon$-machine available to him, the agent may possess complete or partial information. Let us denote the ``full'' $\epsilon$-machine corresponding to complete information by $\mM^*$. Since the logarithmic loss cannot be reduced below that obtained from $\mM^*$, it makes sense to set
\begin{equation}
L(\mM^*)=0
\label{eq:zero-loss}
\end{equation}
and interpret $L(\mM)$ for any other $\epsilon$-machine as the expected logarithmic loss in excess of $h_{\mu}(\mM^*)$. One can easily show that the expected logarithmic prediction loss for a system functioning according to the ``full'' $\epsilon$-machine $\mM^*$ but predicted optimally with the help of another $\epsilon$-machine $\mM$ is equal to the relative entropy rate $h_{\mu}(\mM|\mM^*)$. Using the adapted loss definition (\ref{eq:zero-loss}), we have to subtract the minimum (non-removable) logarithmic expected loss $h_{\mu}(\mM^*)$ to arrive at the loss corresponding to $\mM$. Taking into account the relationship (\ref{eq:hrel}), we arrive at
\begin{equation}
L(\mM)=h_{\mu}(\mM|\mM^*)-h_{\mu}(\mM^*)=\sum_{S\in \mM^*}P_{\mM^*}(S)KL(P_{\mM^*}(S)||P_{\mM}(S)).
\label{eq:L(M)}
\end{equation}

The information quasi-quantity corresponding to the $\epsilon$-machine $\mM$ can be defined, in the absence of any information about possible sources and their knowledge structure, with the help of the most symmetric model for the latter -- uniform and isotropic. This leads, as was mentioned earlier, to the Shannon entropy for the steady state distribution of the corresponding $\epsilon$-machine which is nothing else but the statistical complexity of $\mM$:
\begin{equation}
I(\mM)=-\sum_{S\in \mM}P(S)\log P(S)=C_{\mu}(\mM).
\label{eq:I(M)}
\end{equation}
In particular, if the information available to the agent is complete, i.e. $\mM=\mM^*$, then the information quasi-quantity (\ref{eq:I(M)}) coincides with the system statistical complexity (\ref{eq:C_mu}) (and the loss the agent can achieve vanishes).

Let us now find (or estimate) the quasi-information complexity for the prediction task for the original example. The general definition (\ref{eq:Cgamma}) applied for this case reads
\begin{equation}
C_{\gamma}=\min_{\{\mM : L(\mM)\le \gamma L_0\}} I(\mM),
\label{eq:Cgamma-eps}
\end{equation}
since the role of information $\mI$ is played by various $\epsilon$-machines $\mM$. The original loss $L_0$ here should be identified with the expected logarithmic prediction loss that can be obtained without any knowledge of the underlying statistical mechanism, i.e. using a partial $\epsilon$-machine with a single state. The minimum in (\ref{eq:Cgamma-eps}) has to be taken over all possible $\epsilon$-machines $\mM$ (and corresponding maps $g$: $\mM^*\rightarrow \mM$). Since the full $\epsilon$-machine $\mM^*$ consists of {\it causal} states, it is sufficient to consider partial $\epsilon$-machines with states corresponding to subsets of causal states of $\mM^*$. Specifically, let $\kappa=\{\mS_1,\dotsc, \mS_m\}$ be a {\it transition-consistent} partition of the set $\mS$ of states of the $\epsilon$-machine $\mM^*$, i.e. such a partition that symbols output by any transition between two states of $\mM^*$ are same for all states in the same subset in the partition. An $\epsilon$-machine $\mM$ is obtained from any transition-consistent partition $\kappa$ of the set of states of $\mM^*$ by associating a state $S_j$ of $\mM$ with the subset $\mS_j$ of $\kappa$ for $j=1,\dotsc, m$. The transition probability $P(S_j\rightarrow S_k)$ for $\mM$ is then calculated as $\Pr(S_j\rightarrow S_k)=\sum_{S'\in \mS_j} P(S')\Pr(S'\rightarrow S'')$,
where $S''$ is the state \footnote{It is easy to see that such state $S''$ is unique, due to deterministic property of $\epsilon$-machine $\mM$.} in subset $\mS_k$ with a nonzero transition probability from $S'$.
We denote the resulting $\epsilon$-machine by $\mM(\kappa)$.
Given this construction, the relation (\ref{eq:Cgamma-eps}) can be rewritten as
\begin{equation}
C_{\gamma}=\min_{\{\kappa : L(\mM(\kappa))\le \gamma L_0\}} I(\mM(\kappa)),
\label{eq:Cgamma-eps-k}
\end{equation}
where minimization is over all transition-consistent partitions of the state set of the full $\epsilon$-machine $\mM^*$. 

\begin{figure}
\includegraphics[scale=0.8]{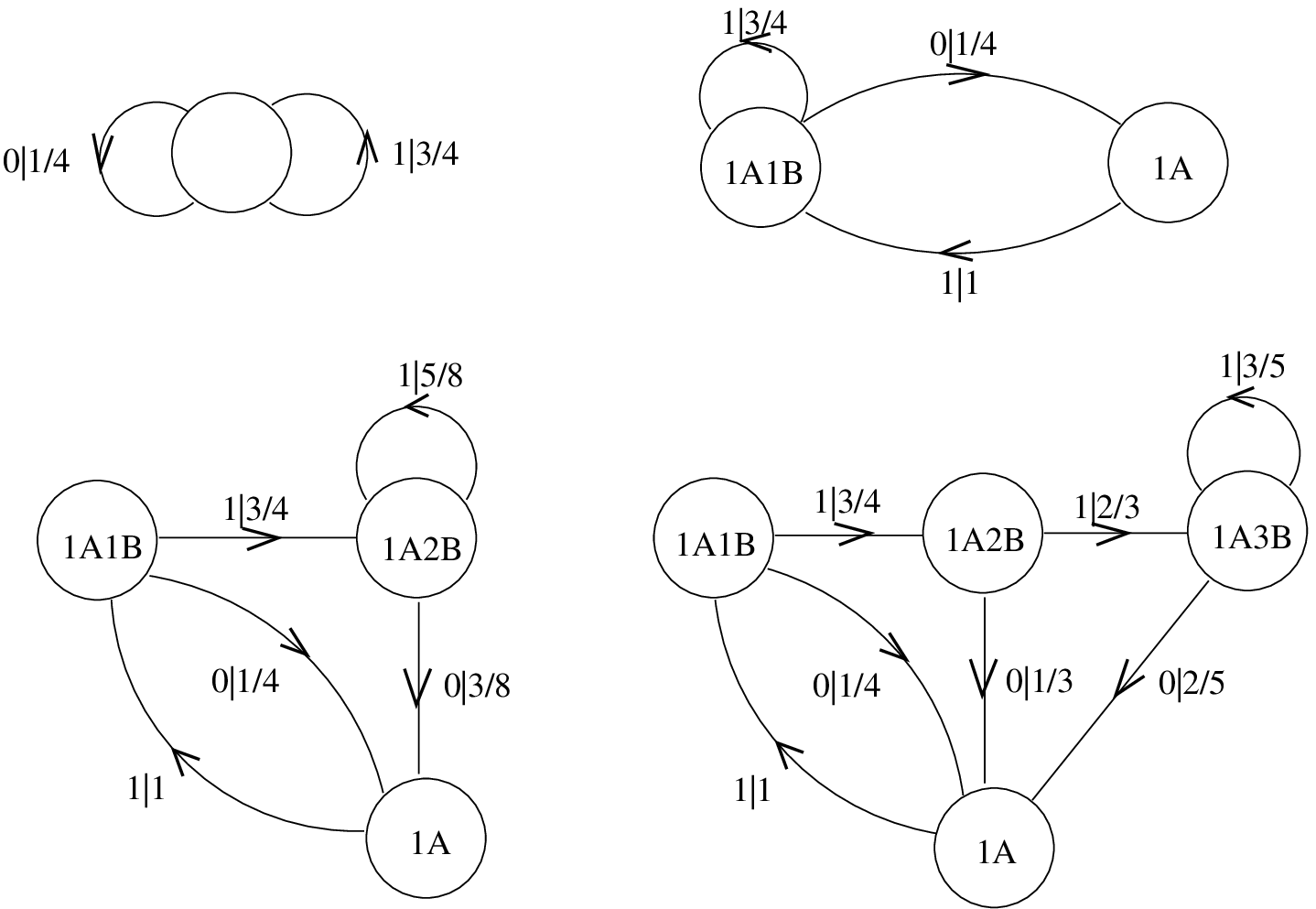}
\caption{\label{f:eps-mach-appr} Sequential approximations to the exact $\epsilon$-machine of Example~A.}
\end{figure}

\begin{figure}
\includegraphics[scale=0.4]{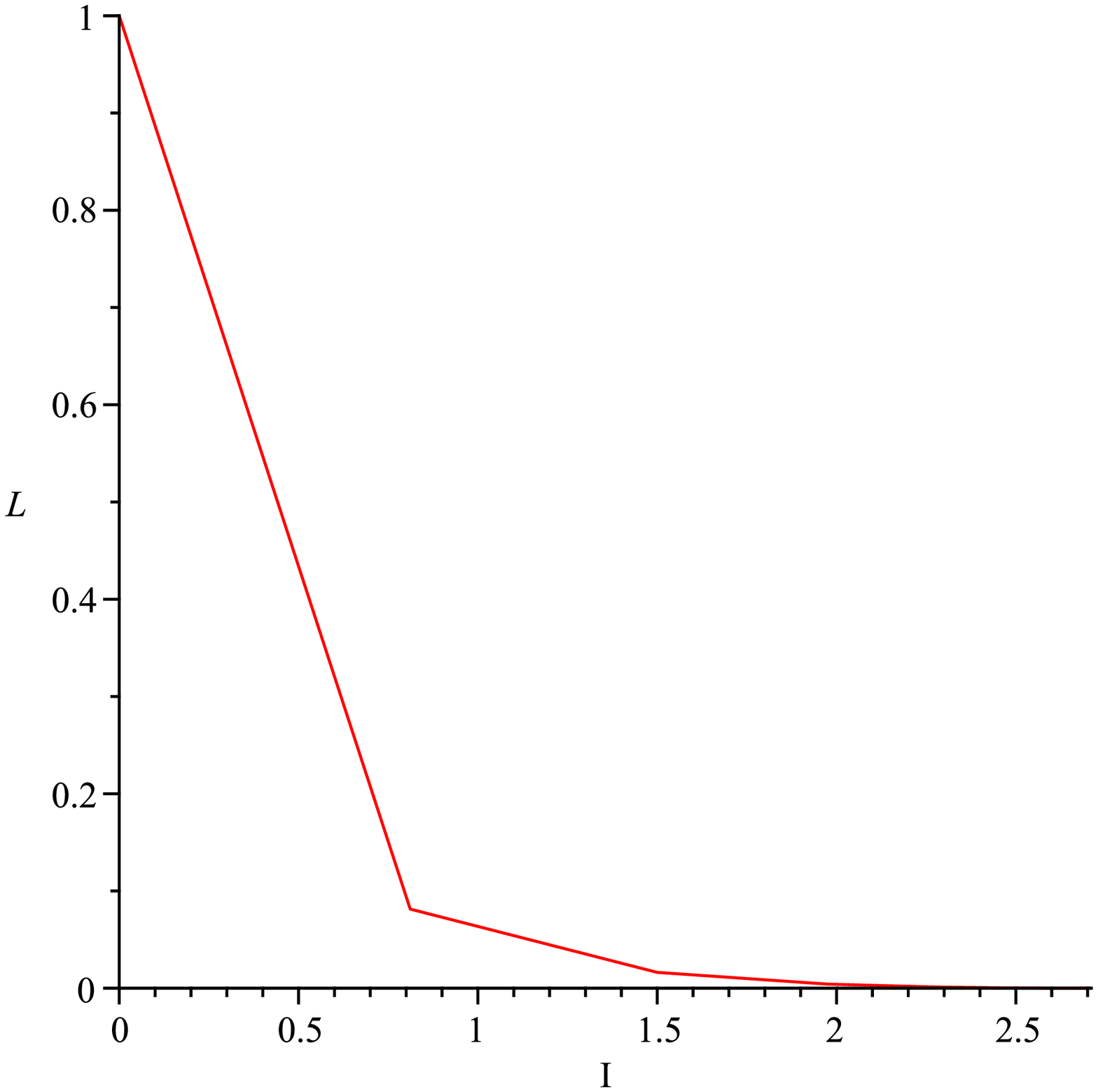}
\caption{\label{f:loss-SGC} Loss vs. information quasi-quantity for Example~A. }
\end{figure}

Finding the exact minimum in (\ref{eq:Cgamma-eps-k}) appears to be a rather difficult problem  since the number of transition-consistent partitions of the state set of  $\mM^*$ is in general exponential in the number of casual states in $\mM^*$. Therefore here we settle for some reasonable choices of ``partial'' information $\mM$ instead. This way, clearly, we will generally obtain upper bounds on the exact values of $C_{\gamma}$. Fig.~\ref{f:eps-mach-appr} shows several such ``partial'' $\epsilon$-machines that obtain naturally in the process of $\epsilon$-machine reconstruction from sample outputs of the system.  Let us denote the ``partial'' $\epsilon$-machines with $k+1$ nodes by $\mM_k$. The map $g$: $\mM^*\rightarrow \mM_k$ for this case has the form $g(1AjB)=1AjB$ for $j<k$ and $g(1AjB)=1AkB$ for $j\ge k$.  The values of relative entropy and statistical complexity of these  $\epsilon$-machines are shown in Table~\ref{t:ex1}. The dependence of the loss on the information quasi-quantity can be found from the data of Table~\ref{t:ex1} by a simple rescaling (so that $L(\mM_1)=1$). The results are shown in Fig.~\ref{f:loss-SGC}. The quasi-information complexity $C_{\gamma}$ for different values of the parameter $\gamma$ can be read directly from the plot. Some of the resulting values are shown in Table~\ref{t:ex1-C}.

\begin{table}
\caption{\label{t:ex1} Relative entropy rates and statistical complexities for sequential approximations to the exact $\epsilon$-machine for Example~A.}
\begin{ruledtabular}
 \begin{tabular}{llllll}
        $k$ & 0 & 1 & 2 & 3 & $\infty$ \\
        $h_{\mu}(\mM_k|\mM^*)$ & 0.811 & 0.708 & 0.683 & 0.679 & 0.678 \\
        $C_{\mu}(\mM_k)$ & 0 & 1.47 & 1.97 & 2.28 & 2.71 \\
\end{tabular}
\end{ruledtabular}
\end{table}

\begin{table}
\caption{\label{t:ex1-C} Quasi-information complexity $C_{\gamma}$ for some values of $\gamma$ for Example~A.}
\begin{ruledtabular}
\begin{tabular}{llllll}
        $\gamma$ & 0.2 & 0.1 & 0.05 & 0.01 & 0 \\
        $C_{\gamma}$ & 0.8 & 1.2 & 1.4 & 1.6 & 2.7 \\
\end{tabular}
\end{ruledtabular}
\end{table}

\begin{figure}
\includegraphics[scale=0.8]{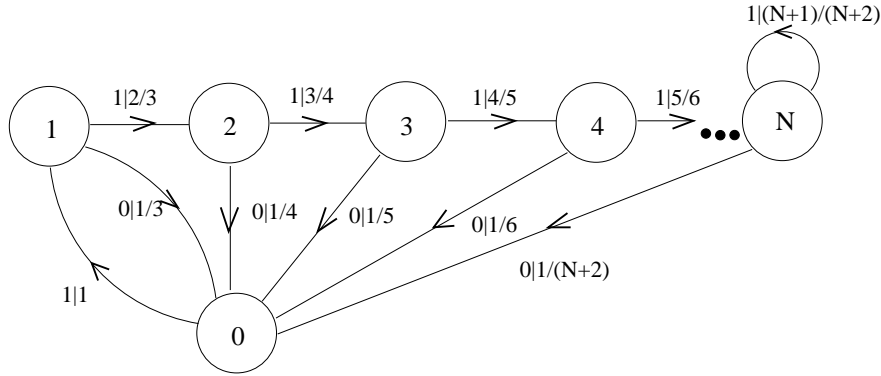}
\caption{\label{f:eps-mach-exact-mod} Exact $\epsilon$-machine for Example~B.}
\end{figure}

\begin{figure}
\includegraphics[scale=0.8]{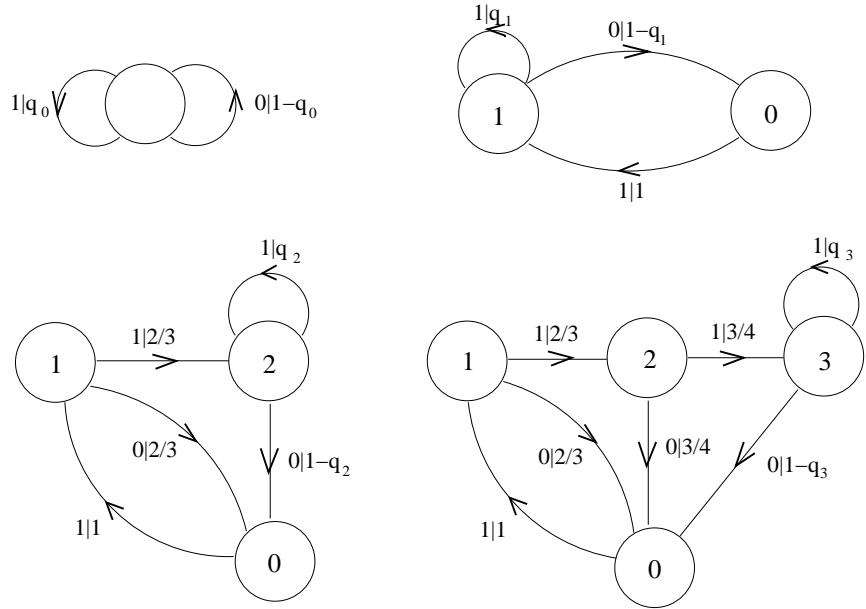}
\caption{\label{f:eps-mach-appr-mod} Sequential approximations to the exact $\epsilon$-machine of Example~B.}
\end{figure}

To consider a different example (called Example~B in what follows), let us keep the topology of the exact $\epsilon$-machine for the example just discussed but change the transition probabilities so that the steady state probabilities do not decay as fast ``down the line'' (see Fig.~\ref{f:eps-mach-exact-mod}). Specifically, let the transition probabilities be $\Pr(i\rightarrow i+1)=\frac{i+1}{i+2}$, $\Pr(i\rightarrow 0)=\frac{1}{i+2}$ for $i=1,\dotsc N-1$, $\Pr(N\rightarrow N)=\frac{N+1}{N+2}$ and $\Pr(0\rightarrow 1)=1$. Just like in Example~A, a transition from a state to another state with a numerically higher label or to itself is accompanied by an output of 1, and a transition from any state to state 0 outputs 0. Note that the number of states for this $\epsilon$-machine has to be finite (but can be made arbitrarily large) in order for steady state probabilities to be positive.  One can easily see that the steady state probabilities for this $\epsilon$-machine are $P(0)=\frac{1}{2(\Psi(N+2)+\gamma_E-1/2)}$ and $P(i)=\frac{2}{i+1}P(0)$, where $\Psi(x)=\frac{\Gamma'(x)}{\Gamma(x)}$ is the digamma function and $\gamma_E\approx 0.577$ is the Euler constant. Let the ``partial'' $\epsilon$-machines representing incomplete information for this example have the same form as in the original example (see Fig.~\ref{f:eps-mach-appr-mod}).

\begin{figure}
\includegraphics[scale=0.4]{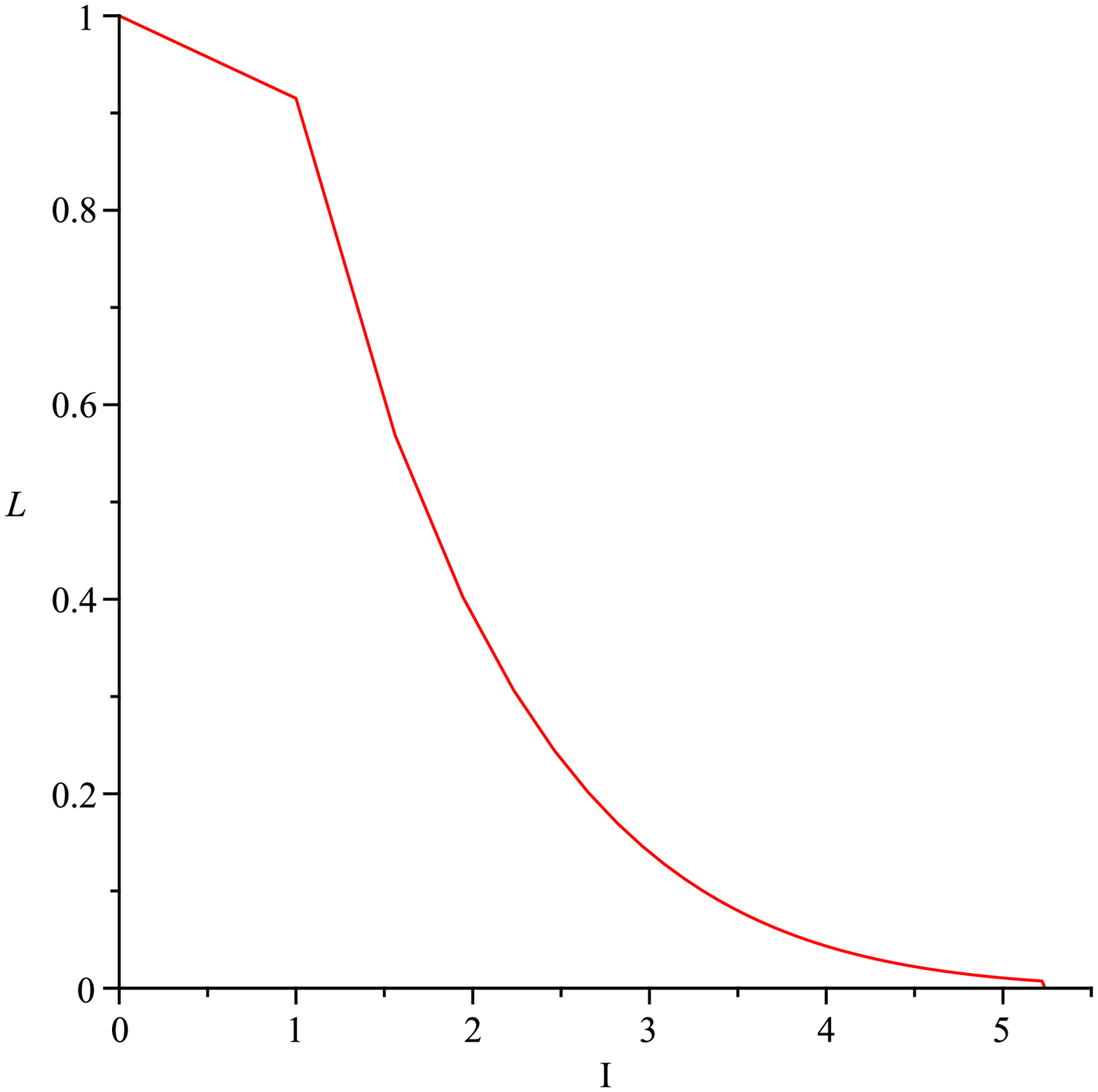}
\caption{\label{f:loss-SGC1} Loss vs. information quasi-quantity for Example~B. }
\end{figure}

The probabilities shown in Fig.~\ref{f:eps-mach-appr-mod} have the following values:
$$q_0=\frac{1}{2}\cdot \frac{1+2(\Psi(N+3)-\Psi(3))}{1+\Psi(N+2)-\Psi(3)} $$
and
$$q_k=\frac{\Psi(N+3)-\Psi(k+2)}{\Psi(N+2)-\Psi(k+1)} $$
for $k=1,2,\dotsc, N$. The steady state probabilities for the ``partial'' $\epsilon$-machines can be computed in the same way as those for the exact one, and thus the information quasi-quantities equal to the corresponding Shannon entropies. On the other hand, the value of loss is found in a straightforward fashion using the general expression (\ref{eq:L(M)}) and taking into account that the map $g$: $\mM^*\rightarrow \mM_k$ in this case has the form $g(j)=j$ if $j<k$ and $g(j)=k$ if $j\ge k$, for the ``partial'' $\epsilon$-machine $\mM_k$ with $k+1$ nodes. The resulting dependence of loss (in the convention $L_0=L(\mM_0)=1$ and setting $N=1000$) on the information quasi-quantity is shown in Fig.~\ref{f:loss-SGC1}. The values of the quasi-information complexity $C_{\gamma}$ for different values of $\gamma$ can be found directly for this dependence. Some of these values are shown in Table~\ref{t:ex2-C}.

\begin{table}
\caption{\label{t:ex2-C} Quasi-information complexity $C_{\gamma}$ for some values of $\gamma$ for Example~B.}
\begin{ruledtabular}
\begin{tabular}{llllll}
        $\gamma$ & 0.2 & 0.1 & 0.05 & 0.01 & 0 \\
        $C_{\gamma}$ & 2.7 & 3.3 & 3.9 & 5.0 & 7.8 \\
\end{tabular}
\end{ruledtabular}
\end{table}

\begin{figure}
\includegraphics[scale=0.8]{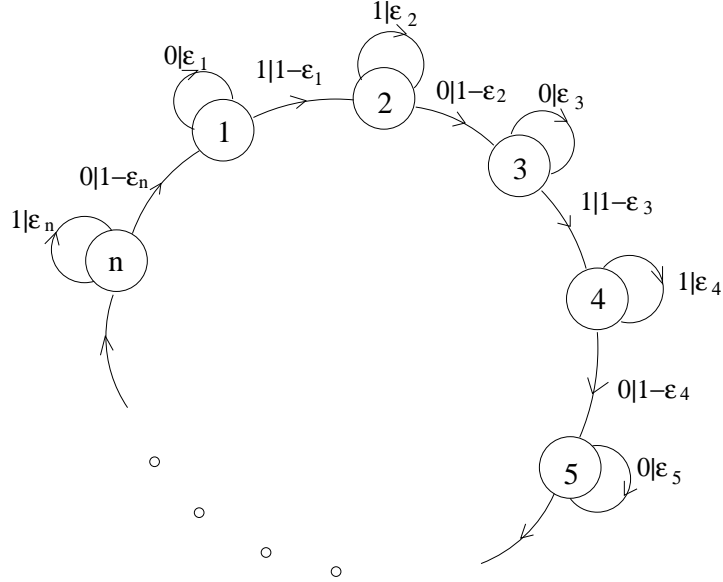}
\caption{\label{f:eps-mach-circ} Exact ``circular'' $\epsilon$-machine for Example~C.}
\end{figure}

Now, consider an $\epsilon$-machine shown in Fig.~\ref{f:eps-mach-circ}. We will refer to this example as Example~C. As can be seen from the figure, this $\epsilon$-machine has $N$ states where $N$ is a (large) even number, and the transition probabilities have the form $\Pr(i\rightarrow i+1)=\epsilon_i$, $\Pr(i\rightarrow i)=1-\epsilon_i$ for $i=1,\dotsc, N-1$, $\Pr(N\rightarrow 1)=\epsilon_N$, $\Pr(N\rightarrow N)=1-\epsilon_N$. Any transition to an even numbered state (either from an odd numbered state or from itself) outputs 1, and any transition to an odd numbered state outputs 0. It is easy to see that the steady state probabilities for this system have the form $P(i)=\frac{1-\epsilon_1}{1-\epsilon_i}P(1)$ for $i=2,\dotsc, N$. If one sets $\epsilon_i=\frac{1}{bi}$ where $b>1$ is a real number, then a simple computation shows that $P(i)=B\frac{bi}{bi-1}$, $i=1,\dotsc, N$, where $B=\left(N+\frac{1}{b}(\Psi(N+1-\frac{1}{b})-\Psi(1-\frac{1}{b}))\right)^{-1}$.

\begin{figure}
\includegraphics[scale=0.8]{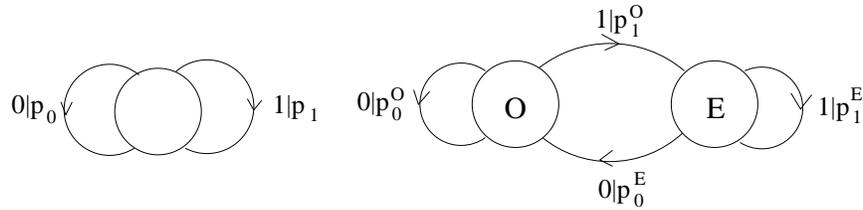}
\caption{\label{f:eps-mach-circ-appr} First two approximations, $\mM_0$ and $\mM_1$, to the ``circular'' $\epsilon$-machine of Example~C.}
\end{figure}

Let the incomplete information about the system be represented by the $\epsilon$-machines shown in Fig.~\ref{f:eps-mach-circ-appr}. The first of these is obtained, just like in the original example, by simply estimating the overall probabilities of 1 and 0 in the output assuming independence of output symbols (thus an absence of any ``information processing'' by the system). A simple calculation gives the following values for the  probability $p_1$:
$$p_1=\frac{1}{2}\cdot \frac{N+\frac{1}{b}\left(\Psi(\frac{1}{2}N+1-\frac{1}{2b})-\Psi(1-\frac{1}{2b}) \right)}{N+\frac{1}{b}\left(\Psi(N+1-\frac{1}{b})-\Psi(1-\frac{1}{b}) \right)}, $$
and $p_0=1-p_1$. The second $\epsilon$-machine in Fig.~\ref{f:eps-mach-circ-appr} is obtained by mapping all odd numbered states of the exact $\epsilon$-machine into the state $O$ and all even numbered states -- into the state $E$. The transition probabilities between states $O$ and $E$ are as shown in Fig.~\ref{f:eps-mach-circ-appr} where
$$p_0^O=\frac{\Psi(\frac{1}{2}N+1-\frac{b+1}{2b})-\Psi(1-\frac{b+1}{2b})}
{Nb+\Psi(\frac{1}{2}N+1-\frac{b+1}{2b})-\Psi(1-\frac{b+1}{2b})}, $$
$$p_1^E=\frac{\Psi(\frac{1}{2}N+1-\frac{1}{2b})-\Psi(1-\frac{1}{2b})}
{Nb+\Psi(\frac{1}{2}N+1-\frac{1}{2b})-\Psi(1-\frac{1}{2b})}, $$
and $p_1^O=1-p_0^O$, $p_0^E=1-p_1^E$, respectively. Any transition into state $O$ is accompanied by an output of 0, and any transition into state $E$ -- by an output of 1.

If one sets the parameter $b$ equal to 3 and $N=1000$, the statistical complexity of the system calculates to be equal to 9.97. On the other hand, for the same values of parameters $b$ and $N$, the loss corresponding to the two ``partial'' $\epsilon$-machines shown in Fig.~\ref{f:eps-mach-circ-appr} is $L(\mM_0)=0.98$ and $L(\mM_1)=0.008$, with their information quasi-quantities (Shannon entropies) being equal to 0 and 1.0, respectively. It follows that, $C_{0.008}=1.0$ for this system (or, rather, that 1.0 is a upper bound on $C_{0.008}$ as explained earlier).

We can now make some interesting observations.
\begin{itemize}
\item Comparing Examples A and B, we notice that the ``full'' quasi-information complexity $C_0$ (which is equal to statistical complexity) in Example~B is significantly higher (7.8 vs 2.7) reflecting higher steady state probabilities of higher numbered states for Example~B.
\item The quasi-information complexity values for $\gamma>0$ are also higher in the absolute sense but are about the same in the relative (measured as fractions of $C_0$) sense for Example~B, compared to Example~A. The latter observation can likely be explained by the two $\epsilon$-machines sharing the same topology.
\item If we now compare Examples A and C, we can see that the statistical complexity of the system in Example~C is significantly higher than that in Example~A (10.0 vs 2.7), again reflecting higher steady state probabilities of larger number of states in Example~C.
\item On the other hand, the quantity $C_{0.01}$ is actually {\it lower} for the ``circular'' system in Example~C (1.0 vs. 1.6). Informally speaking, this means that if  a relative 1\% difference in prediction quality is immaterial to the agent, the system in Example~C would appear less complex than that in Example~A.
\item In this particular case, the latter phenomenon has a simple explanation. Indeed, if the agent predicts the next bit based on the hypothesis of i.i.d. distribution (which corresponds to the agent using the simplest $\epsilon$-machine $\mM_0$ possessing no ``structure'') the resulting error is a lot larger than that afforded by the knowledge of the `full'' $\epsilon$-machine (complete information). In fact, the bits appear to be almost ``unpredictable'' (as the estimated probability of 1 is close to 0.5).  The situation changes drastically if the agent is able to
    get access to $\mM_1$ with two causal states, i.e. able to realize that in this system 0 is followed by 1 and vice versa with high probability. The knowledge of $\mM_1$ allows the agent to predict nearly as well as if he had access to the full $\epsilon$-machine $\mM^*$ with 1000 causal states. Put slightly differently, once the realization is made that 0 and 1 symbols ``mostly'' alternate in this system, the rest of information just introduces small corrections to this dominant behavior.
\item Summarizing, one can say that to an agent not concerned (for whatever reason) with small (1\% or less) ``residual'' losses, the system from Example~C would appear as genuinely less complex than that in Example~A, in spite of a much larger statistical complexity in Example~C.
\end{itemize}
Informally speaking, an agent observing the system output in Example~C would quickly observe that 0's and 1's mostly alternate but sometimes shorts strings of the same symbol are observed. This observation would let the agent conclude that the system can be reasonably accurately described by an $\epsilon$-machine of the form $\mM_1$ presented earlier. It would take the agent, generally speaking, a lot more effort to discern finer details: namely that length of these short string of the same symbol tend to change just a little as time passes. The discovery of these fine details though would improve the agent's predictive ability only very slightly. One could say that, for this particular system, it is relatively easy to understand its behavior ``almost fully'' but a lot more difficult to fill in the remaining small details. The initially very steep $C_{\gamma}$ curve that flattens out drastically reflects that. 

\section{\label{s:conclusion}Conclusion}
The main contributions of this article can be briefly summarized as follows.
\begin{itemize}
\item The notion of information-related complexity was extended from systems described by stochastic processes to more general tasks (performed by a rational agent), including optimization and decision making problems with uncertainty about input data.
\item A general problem oriented definition of complexity was proposed and it was shown that it reduces to a generalization of statistical complexity in the case the system output is described by a discrete-time stochastic process and the agent's task is that of prediction .
\item The degree of task completion was explicitly considered and it was argued that the information-related complexity should be properly understood relative to the extent of task completion, or, equivalently, properly defined loss. In other words, the information-related complexity for the given task is inherently a real-valued function, and not just a real number.
\item It was argued, based on results of \cite{part1,part2}, that the minimum information quantity necessary to complete the task to the given degree should be really interpreted as appropriate quasi-quantity that takes into account the degree of difficulty of obtaining the corresponding information in the given environment. The traditional information quantity measures (such as Shannon entropy) obtain from the appropriate quasi-quantity in the particular case of maximum symmetry, when either all different ``pieces'' of information are identical (with respect to the degree of difficulty of their acquisition by the agent) or their differences are unknown.
\end{itemize}

Let's discuss some of these points in an informal way. 
A more complex system is the one that ``takes a lot of work'' to fully understand for a rational agent, and, moreover, a complex system still takes a lot of work to understand ``almost fully''. On the contrary, a system that is ``easy'' to understand to a large degree but very hard to ``get the rest'' (a system with for which the $C_{\gamma}$ curve is initially very steep but flattens out at close to zero loss) would appear to be of low complexity to all but ``very discerning'' agents. One might ask: what about the true ``objective'' complexity of a system that's independent (being objective) on all and any agents. We believe that this question might be more difficult and deeper than it appears. One possible answer is that it depends on the point of view and the level of inquiry. The same system can be complex from one point of view and much less so from another. Additionally, it can be complex at high level of detail but much less complex at ``lower resolution''.
For instance, in natural sciences, a system's ability to ``innovate'', or spontaneously create new quality could be due to various levels of its structure detail. Hence, its appropriate complexity would also depend on that. Generally speaking, it is our opinion that the science of complexity still has more questions than answers, and it would be of use, for the sake of clarity, to make some implicitly used or assumed aspects of the overall problem -- explicit. This is one of the main goals of the present article.


%

\end{document}